\documentclass[12pt]{article}
\usepackage{amsfonts,amssymb}

\def\p{ \partial }

\def\bq{ \begin{equation} }
\def\eq{ \end{equation} }
\def\ben{ \begin{eqnarray} }
\def\en{ \end{eqnarray} }

\def\frac#1#2{{#1\over #2}}

\def\on#1#2{\mathop{\vbox{\ialign{##\crcr\noalign{\kern2pt}
$\scriptstyle{#2}$\crcr\noalign{\kern2pt\nointerlineskip}
\kern-2pt$\hfil\displaystyle{#1}\hfil$\crcr}}}\limits}

\begin{document}

\baselineskip=15pt
\vspace{1cm} \centerline{{\LARGE \textbf {Non-homogeneous 
systems of hydrodynamic 
 }}}
\vspace{0.3cm} \centerline{{\LARGE \textbf {type possessing Lax representations
 }}}

\vskip1cm \hfill
\begin{minipage}{13.5cm}
\baselineskip=15pt {\bf A.V. Odesskii ${}^{1} $,  V.V. Sokolov
${}^{2}$}
\\ [2ex] {\footnotesize
${}^{1}$  Brock University (Canada)
\\
${}^{2}$  L.D. Landau Institute for Theoretical Physics (Russia)
\\}
\vskip1cm{\bf Abstract.} We consider  $1+1$ - dimensional non-homogeneous systems of hydrodynamic type that possess  Lax 
representations with movable singularities. We present a construction, which provides a wide class of examples of such systems with arbitrary number of components. 
In the two-component case a classification is given. 

\end{minipage}

\vglue1cm \textbf{Address}: L.D. Landau Institute for Theoretical
Physics of Russian Academy of Sciences, Kosygina 2, 119334,
Moscow, Russia

\textbf{E-mail}: aodesski@brocku.ca, sokolov@itp.ac.ru
\newpage

\section{Introduction}

The integrability theory for $1+1$ - dimensional homogeneous systems of hydrodynamic type of the form 
\begin{equation}   \label{genhom}
u_{i,t}=\sum_{j=1}^n
a_{ij}({\bf u})\,u_{j, x}, \qquad i=1,...,n,
\end{equation}
where ${\bf u}=(u_{1},\dots, u_{n})$ was developed in \cite{dubnov, tsar}. In the case of $1+2$ - dimensional homogeneous systems  
a definition of integrability based on the existence of hydrodynamic reductions was proposed in \cite{ferhus1}. For examples and a classification of integrable cases see \cite{odsok2} and references therein. 

In contrast to the homogeneous case,  there is no satisfactory criteria of integrability for non-homogeneous hydrodynamic type systems of the form
\begin{equation}   \label{geninhom}
u_{i,t}=\sum_{j=1}^n
a_{ij}({\bf u})\,u_{j, x}+b_i({\bf u}), \qquad i=1,...,n.
\end{equation}
On the other hand, the following example 
\begin{equation}   \label{gibts}
u_t=v u_x+\frac{1}{v-u}, \qquad v_t=u v_x+\frac{1}{u-v}
 \end{equation}
found in  \cite{Gibt1} indicates the existence of integrable systems (\ref{geninhom}) 
having properties unusual for $1+1$-dimensional integrable models. In particular, system (\ref{gibts}) has only a few local infinitesimal symmetries and therefore the symmetry approach to classification of 
integrable $1+1$-dimensional systems \cite{mss} is not applicable for systems similar to (\ref{gibts}). It turns out that this system possesses infinite non-commutative hierarchy of nonlocal symmetries explicitly depending on $x$ and $t$. These symmetries look like symmetries of the so called Gibbons-Tsarev systems found in \cite{odsok3}.

Notice that system (\ref{gibts}) was also derived in another context, in \cite{ferfor, marsok} where conditions
for two quadratic Hamiltonians to be in involution were studied.

It was shown in \cite{Gibt2} that the general solution of (\ref{gibts}) can be described in terms 
of a conformal mapping of a slit domain to the upper half plane (cf. \cite{loew,tt}). A class of special solutions was constructed in \cite{kokkor}. All these results are related to the existence of the following Lax representation 
\begin{equation}   \label{laxgibts}
\Psi_t=\frac{\lambda-u-v}{(\lambda-u)(\lambda-v)} \Psi_{\lambda}, \qquad \Psi_x=-\frac{1}{(\lambda-u)(\lambda-v)} \Psi_{\lambda}
 \end{equation}
for (\ref{gibts}). Here  $\lambda$ is a spectral parameter. 

The system (\ref{gibts})  has the conservation law \quad 
$
(u+v)_t=(u v)_x.
$
Introducing the function $Z$ such that $\quad Z_x=u+v, \, Z_t=u v,$ we arrive at
$$
Z_{tt}-Z_x Z_{xt}+Z_t Z_{xx}=1.
$$
Anzats for known explicit solutions of this equation are:
$$
Z=x^3+a_2(t) x^2+a_1(t) x+a_0(t) 
$$
and 
$$
Z=a_1(t) e^x+a_2(t) e^{-x}+a_0(t).
$$
In the second case \cite{kaptsov} we have $a_0'=t-2 a_1 a_2$ and
$$
a_1''-2 a_2 a_1^2+x a_1=0, \qquad a_2''-2 a_1 a_2^2+x a_2=0.
$$
The latter system can be reduced to the Painleve II equation. The ODE system for $a_i$ in the first case leads to the Painleve I.

In this paper we consider systems (\ref{geninhom}) having Lax representations of the form
\begin{equation} \label{lax} 
\Psi_x=f ({\bf u}, \lambda) \Psi_{\lambda}, \qquad \Psi_t=g ({\bf u}, \lambda) \Psi_{\lambda}.
\end{equation} 
For brevity, we call such systems {\it integrable}. Lax pairs similar to (\ref{lax}) were considered in \cite{bzm}. Note that Lax representations with $\lambda$-derivatives of the $\Psi$-function 
are typical for Painleve type equations. Possibly the technique developed in \cite{mansan} for solving of dispersionless systems can be adopted to systems with such Lax pairs.

To generalize example (\ref{gibts}),  (\ref{laxgibts}) we assume (cf. \cite{odsok1}) that both functions $f$ and $g$ in (\ref{lax})  
have simple  poles at $\lambda=u_1,...,u_n$. In general, we do not restrict ourselves to functions $f,g$ rational in $\lambda$ but all examples in this paper are rational. 

In Section 2 we show that for any system (\ref{geninhom}) with a Lax representation of the kind described above the matrix   ${\bf a}=(a_{ij}(\bf u) )$ is diagonal and weakly non-linear \cite{fer}. The latter property means 
that the functions $a_{ii}(\bf u)$ do not depend on $u_i$.
We establish a partial separation of variables for functions $f$ and $g$ in (\ref{lax}) and write down a general functional equation describing these functions.  

In Section 3 we construct a large class of examples of integrable systems of the form (\ref{geninhom}) with arbitrary $n$. These systems depend on parameters. The main continuous parameters are the roots $\lambda_i$ of a polynomial 
with constant coefficients. To each $\lambda_i$ a non-negative integer $k_i$ is attached. In Section 3.1 the case is considered when all the  $k_i$ are zero. In Section 3.2 we describe a limit of formulas from Section 3.1 when some of the $\lambda_i$ coincide. In Section 3.3 we generalize results obtained in  
Section 3.1 to the case of arbitrary $k_i.$ 
 
Section 4 is devoted to the case $n=2$. We find all possible pairs of functions  $f$ and $g$ appearing in the Lax representation (\ref{lax}).  Even for $n=2$ the functional equation for functions  $f$ and $g$ is highly non-trivial. 
It turns out that it is almost equivalent to a functional equation describing the functions $F,H$ in the fieldless Gibbons-Tsarev type systems 
\begin{equation} \label{GTsys}
\partial_i \xi_{j}=F(\xi_{i},\xi_{j} )\, \partial_i u ,
\qquad
\partial_i \partial_{j} u = H(\xi_{i},\xi_{j} ) \,  \partial_i u \partial_j
u,\qquad i\ne j, \qquad i,j=1,...,N.
\end{equation}
Here $u, \xi_{i} $ are functions of variables $r^{1},\dots,r^{N},$    and $\partial_i=\frac{\partial}{\partial_{r^i}}.$  Such systems play a crucial role  \cite{ferhus1, odsok2} in the integrability theory 
for $1+2$-dimensional homogeneous hydrodynamic type systems. As an additional result we find new examples of Gibbons-Tsarev type systems (\ref{GTsys}). From our classification it follows  that in the case $n=2$ 
all systems with Lax representations (\ref{lax}) can be constructed by the approach developed in Section 3. 

In Section 5 we present some explicit examples of integrable systems of the form (\ref{geninhom}) in the case $n=3$.

\section{Lax pairs with movable singularities}

Our goal is a generalization of example (\ref{gibts}),  (\ref{laxgibts}). In this example both functions $f$ and $g$ in (\ref{lax})  
have simple  poles at $\lambda=u$ and $\lambda=v$. In the general case let us assume (cf. \cite{odsok1}) that the functions $f$ and $g$ in (\ref{lax})  
have simple  poles at $\lambda=u_1,...,u_n$.

Computing $(\Psi_x)_t$ and $(\Psi_t)_x$ by virtue of (\ref{geninhom}) and (\ref{lax}) and equating coefficients at $u_{j, x},~j=1,...,n,$ we obtain 
the following compatibility conditions:
\begin{equation}   \label{comp1}
 \frac{\p g}{\p u_i}=\sum_{j=1}^n a_{ji} \frac{\p f}{\p u_j} ,~~~i=1,...,n,
\end{equation}
\begin{equation}   \label{comp2}
 \sum_{i=1}^nb_i \frac{\p f}{\p u_i}+f \frac{\p g}{\p \lambda}-g \frac{\p f}{\p \lambda}=0.
\end{equation}
Computing the singular part of (\ref{comp1}) at $\lambda=u_j,$ where $j\ne i,$ we 
obtain $a_{ji}=\delta_{i,j}a_i$. We assume that the functions $a_i$   are pairwise distinct. Now our system (\ref{geninhom}) reads:  
\begin{equation}   \label{diaginhom}
u_{i,t}=a_{i}({\bf u})\,u_{i, x}+b_i({\bf u}), \qquad i=1,...,n.
\end{equation}
The compatibility condition (\ref{comp1}) becomes
\begin{equation}   \label{comp3}
 \frac{\p g}{\p u_i}=a_i \frac{\p f}{\p u_i},~~~i=1,...,n.
\end{equation}
Let $$f=\frac{\phi_i}{\lambda-u_i}+O(\lambda-u_i),  \qquad g=\frac{\psi_i}{\lambda-u_i}+O(\lambda-u_i), \qquad \phi_i\ne 0,\quad \psi_i\ne 0$$ for $\lambda\to u_i$. Computing the singular parts of (\ref{comp3}), 
we obtain 
$$\phi_i=a_i\psi_i,\qquad \frac{\p \phi_{j}}{\p u_i}=a_i \frac{\p \psi_{j}}{\p u_i}$$
for all $i,j=1,...,n$. Substituting the first of these equations into the second one, we obtain $(a_j\psi_j)_{u_i}=a_i\psi_{j,u_i}$. From here it follows  that if $i=j$, then 
$$\frac{\p a_{i}}{\p u_i}=0,$$
otherwise
$$\frac{\p \psi_{j}}{\p u_i}= \frac{1}{a_i-a_j} \frac{\p a_{j}}{\p u_i} \psi_j,\qquad i\ne j.$$
The first of these equations shows that the homogeneous system 
\begin{equation}\label{weak} u_{i,t}=a_i u_{i,x},\qquad i=1,...,n
\end{equation} 
is weakly non-linear. The compatibility conditions for the second equation
means that this homogeneous system is semi-Hamiltonian. Weakly non-linear semi-Hamiltonian systems (\ref{weak}) were studied by E. Ferapontov in \cite{fer}. In particular, he classified 
these systems finding possible coefficients $a_i$ in a closed form in terms of arbitrary functions of one variable. Namely,  for any weakly non-linear semi-Hamiltonian system (\ref{weak}) the coefficient $a_i$  has the form:
\begin{equation} \label{aaa} a_i=\frac{\det\Delta_{n,i}}{\det\Delta_{n-1,i}}.
\end{equation}
Here and in the sequel  
\begin{equation} \label{delta}
\Delta=\left(\begin{array}{ccc}1&...&1\\q_{1,1}(u_1)&...&q_{1,n}(u_n)
\\.........&...&.........\\q_{n-1,1}(u_1)&...&q_{n-1,n}(u_n)
\end{array}\right),\end{equation}
where $q_{i,j}$ are arbitrary functions of one variable. The meaning of the indexes on $\Delta$ in (2.12) is as follows. For any matrix $M$ we denote by $M_{i,j}$ its $i,j-$minor. In other words, $M_{i,j}$ is obtained from $M$ by deleting its $i-$th 
row and $j-$th column. By $M_i$ we denote the matrix obtained from $M$ by deleting its $i-$th row.  

For functions (\ref{aaa}) the general solution of (\ref{comp3}) is given by
\begin{equation}   \label{fgdet}
f=\frac{\det P}{\det\Delta}, \qquad g=\frac{\det Q}{\det\Delta}, 
\end{equation}
where 
$$P=\left(\begin{array}{ccc}h_1(\lambda,u_1)&...&h_n(\lambda,u_n)\\&\Delta_{n-1}\end{array}\right),\qquad Q=
\left(\begin{array}{ccc}h_1(\lambda,u_1)&...&h_n(\lambda,u_n)\\&\Delta_n
\end{array}\right).$$
Here $h_1,...,h_n$ are arbitrary functions of two variables. Notice that $P_n=Q_n$.
According to our assumption each function $h_i(\lambda,u)$ has a simple pole at $\lambda=u$. Substituting these expressions for $f,~g$ into (\ref{comp2}) and computing the 
singular part at $\lambda=u_i$, we get
\begin{equation}   \label{b}
b_i=\frac{\det Q_{n,i}|_{\lambda=u_i}}{\det\Delta_{n-1,i}}, \qquad i=1,...,n. 
\end{equation}
Substituting these into (\ref{comp2}), we obtain a functional equation, which can be written in the following form:
\begin{equation}   \label{funeqgen}
\det \left(\begin{array}{ccc}h_1(\lambda,u_1)_{\lambda}&...&h_n(\lambda,u_n)_{\lambda}\\&Q_n&
\end{array}\right) +\sum_{i=1}^n (-1)^{i-1}h_i(\lambda,u_i)_{u_i}\det Q_{n,i}|_{\lambda=u_i}+
\end{equation}
$$\frac{1}{\det\Delta}\sum_{1\leq k_1\leq n-1,1\leq k_2\leq n}(-1)^{k_1+k_2}\det Q_{n,k_2}|_{\lambda=u_{k_2}}\det 
\left(\begin{array}{ccc}h_1(\lambda,u_1)&...&h_n(\lambda,u_n)\\&\Delta_{k_1+1}\end{array}\right)q^{\prime}_{k_1,k_2}(u_{k_2})=0.$$

{\bf Remark 1.} Expanding numerators of (\ref{fgdet}) by the first row, we get
$$f=h_1(\lambda,u_1)\phi_1+...+h_n(\lambda,u_n)\phi_n, \qquad g=h_1(\lambda,u_1)\psi_1+...+h_n(\lambda,u_n)\psi_n,$$
where 
$$\phi_i=(-1)^{i-1}\frac{\Delta_{n-1,i}}{\Delta_{n-1}}, \qquad \psi_i=(-1)^{i-1}\frac{\Delta_{n,i}}{\Delta_{n}}.$$
Note that according to the Cramer rule, $\phi_i,~\psi_i$ are the solutions of the following system of linear equations:
\begin{equation}   \label{lins}
q_{j,1}\phi_1+...+q_{j,n}\phi_n=\delta_{n-2,j}, \qquad q_{j,1}\psi_1+...+q_{j,n}\psi_n=\delta_{n-1,j},  
\end{equation}
where $j=0,...,n-1$, $q_{0,i}(u)=1$ and $\delta_{i,j}$ is the Kronecker delta.  $\square$

{\bf Remark 2.} For all known examples the functions $h_i(\lambda,u),~q_{j,i}(u)$ don't depend on $i$. We conjecture that this property holds for any solution of (\ref{funeqgen}). This conjecture is proved in the cases $n=2,~3$.  $\square$

\section{A class of integrable systems with arbitrary number of components}

The compatibility condition for (\ref{lax}) reads\footnote{Note that this equation can be written as $\frac{\partial f}{\partial t}-\frac{\partial g}{\partial x}+[f,g]=0,$ 
where $[f,g]=f\frac{\partial g}{\partial \lambda}-g \frac{\partial f}{\partial \lambda}$ is the bracket in an algebra of vector fields. Therefore (\ref{comp}) can be considered as zero 
curvature equation for a connection in an $Diff^1$-bundle, where $Diff^1$ is the group of diffeomorphisms of one-dimensional manifold with coordinate $\lambda$.}
\begin{equation} \label{comp} 
\frac{\partial f}{\partial t}+f\frac{\partial g}{\partial \lambda}=\frac{\partial g}{\partial x}+g \frac{\partial f}{\partial \lambda}.
\end{equation} 
In this section we assume that $f$ and $g$ are rational functions in $\lambda$. In addition, we suppose that both  $f$ and $g$ are   divisible by a polynomial $S(\lambda)$ with constant coefficients. At first glance 
this is not true for our basic example (\ref{laxgibts}). Nevertheless, under the transformation
\begin{equation}\label{frac}
\lambda\to\bar{\lambda}=\frac{a\lambda+b}{c\lambda+d}
\end{equation}
with constant $a,b,c,d$ a common multiplier $(c\lambda+d)^3$ in $f$ and $g$ appears. 

One can construct conservation laws for a system of the form (\ref{geninhom}) using such a Lax 
representation (\ref{lax}) as follows:

{\bf Lemma 1.} Suppose that $f=S(\lambda)f_1, \,\, g=S(\lambda)g_1,$ where  $S(\lambda)$ is a polynomial with constant coefficients and with roots $\lambda_1,...,\lambda_p$. Set 
$\mu_i=f_1|_{\lambda=\lambda_i},~and \nu_i=g_1|_{\lambda=\lambda_i}.$ Then 
\begin{equation}\label{cons}
\frac{\partial \mu_i}{\partial t}=\frac{\partial \nu_i}{\partial x}.
\end{equation}
Moreover, if $\lambda_i$ is a root of multiplicity $k$ and $\displaystyle \mu_{i,j}=\frac{\partial^jf_1}{\partial\lambda^j}|_{\lambda=\lambda_i},~\nu_{i,j}=
\frac{\partial^jg_1}{\partial\lambda^j}|_{\lambda=\lambda_i},$
where $j=0,...,k-1$, then 
\begin{equation}\label{cons1}
\frac{\partial \mu_{i,j}}{\partial t}=\frac{\partial \nu_{i,j}}{\partial x}.
\end{equation}
{\bf Proof.} Substituting our expressions for $f,~g$ into (\ref{comp}), we get 
\begin{equation} \label{comp0}
\frac{\partial f_1}{\partial t}+S(\lambda)\frac{\partial g_1}{\partial \lambda} f_1=\frac{\partial g_1}{\partial x}+S(\lambda)\frac{\partial f_1}{\partial \lambda} g_1.                                                                                
\end{equation}
To complete the proof we set $\lambda=\lambda_i$ here. In the case of multiple roots we compute derivatives of (\ref{comp0}) with respect to  $\lambda$ and then set $\lambda=\lambda_i$.  $\square$

\subsection{Principal series of examples}

Let 
\begin{equation}\label{fgrep}
 f=\frac{S(\lambda)P(\lambda)}{R(\lambda)},\qquad g=\frac{S(\lambda)Q(\lambda)}{R(\lambda)},
\end{equation}
where $S,~P,~Q,~R$ are polynomials in $\lambda$ and all coefficients of $S$ are constant. 
Then equation (\ref{comp}) reads as
\begin{equation} \label{comp4} 
R \frac{\partial P}{\partial t} - P\frac{\partial R}{\partial t}+Q \frac{\partial R}{\partial x}-R\frac{\partial Q}{\partial x}+S (\lambda)\left(P\frac{\partial Q}{\partial \lambda}-Q\frac{\partial P}
{\partial \lambda}\right)=0.
\end{equation} 

{\bf Proposition 1.} Let $$h_i(\lambda,u)=\frac{S(\lambda)}{\lambda-u}, \qquad q_{j,i}(u)=u^j,$$
if $i=1,...,n,~j=1,...,n-m-3$ and 
\begin{equation} \label{qs}
q_{n-m-3+j,i}(u)=\sum_{k=1}^{n+1-m}\frac{c_{k,j}}{u-\lambda_k} 
\end{equation}
if $i=1,...,n,~j=1,...,m+2$. Here $m$ is a fixed integer such that   $0\leq m\leq \frac{n-1}{2}$, $S(\lambda)=(\lambda-\lambda_1)...(\lambda-\lambda_{n+1-m}),$ where   $\lambda_i$ are pairwise distinct constants, 
and $c_{i,j}$ are arbitrary constants such that the matrix $(c_{i,j})$ has rank $m+2.$
Then formulas (\ref{delta}) - (\ref{b})  define a system (\ref{diaginhom}) 
possessing a Lax representation (\ref{lax}).

{\bf Proof.} Expanding the determinants in the numerators of (\ref{fgdet}) by the first row, we get 
\begin{equation} \label{fg0}
 f=S(\lambda)\left(\frac{\phi_1}{\lambda-u_1}+...+\frac{\phi_n}{\lambda-u_n}\right),\qquad g=S(\lambda)\left(\frac{\psi_1}{\lambda-u_1}+...+\frac{\psi_n}{\lambda-u_n}\right),
\end{equation}
where  $\phi_1,...,\phi_n, \psi_1,...,\psi_n$ satisfy the linear system (\ref{lins}) (see Remark 1 in Section 2). The part of this system corresponding to  $j=0,...,n-m-3$ has the form
\begin{equation} \label{uuu} u_1^j\phi_1+...+u_n^j\phi_n=0, \qquad u_1^j\psi_1+...+u_n^j\psi_n=0.
\end{equation}
From these equations it follows  that
\begin{equation} \label{ffgg} f=\frac{S(\lambda)P(\lambda)}{(\lambda-u_1)...(\lambda-u_n)}, \qquad g=\frac{S(\lambda)Q(\lambda)}{(\lambda-u_1)...(\lambda-u_n)},
\end{equation}
where $P,~Q$ are polynomials in $\lambda$ 
of degree $m+1$. Indeed, using expressions (\ref{fg0}) and expanding  $\frac{f}{S}$ and $\frac{g}{S}$ with respect to powers of $\lambda^{-1},$ we get left hand sides of (\ref{uuu}) as coefficients at $\lambda^{-j}$, 
$j=1,...,n-m-2$. 
The left hand side of (\ref{comp4}) is a polynomial in $\lambda$ of degree $n+m+1.$ Equating its coefficients to zero, we obtain $n+m+2$ non-homogeneous differential equations of hydrodynamic type for $u_1,...,u_n$. To obtain a system of the form 
(\ref{geninhom}) we have to prove that there 
exist $m+2$ linear dependence relations between these equations. Let 
\begin{equation} \label{munu}
 \mu_i=\frac{\phi_1}{\lambda_i-u_1}+...+\frac{\phi_n}{\lambda_i-u_n},\qquad \nu_i=\frac{\psi_1}{\lambda_i-u_1}+...+\frac{\psi_n}{\lambda_i-u_n},
\end{equation}
where $~i=1,...,n+1-m.$ 
According to Lemma 1, we have $n+1-m$ differential equations (\ref{cons}). On the other hand, the relations
\begin{equation} \label{lind} 
c_{1,j}\mu_1+...+c_{n+1-m,j}\mu_{n+1-m}=\delta_{m+1,j}, \qquad c_{1,j}\nu_1+...+c_{n+1-m,j}\nu_{n+1-m}=\delta_{m+2,j}                     \end{equation} 
for $j=1,...,m+2$, are fulfilled. 
Indeed, substituting the expressions (\ref{munu}) for $\mu_i,~\nu_i$ into (\ref{lind}), we get the part of the linear system (\ref{lins}) corresponding to $j=n-m-2,...,n-1,$ where the functions $q_{j,i}$ are given by (\ref{qs}). 
Relations (\ref{lind})
give us $m+2$ linear dependence relations between equations (\ref{cons}), namely $$c_{1,j}((\mu_1)_t-(\nu_1)_x)+...+c_{n+1-m,j}((\mu_{n+1-m})_t-(\nu_{n+1-m})_x)=0.$$  Since $(\mu_i)_t-(\nu_i)_x$ is proportional to the 
left hand side of (\ref{comp4}), where $\lambda$ is set to $\lambda_i$, we obtain $m+2$ linear dependences between coefficients of the left hand side of (\ref{comp4}).
Note that $\mu_i$ (resp. $\nu_i$) are given by the formula (\ref{fgdet}) for $f$ (resp. for $g$), where $h_j(\lambda,u_j)$ are replaced by $\frac{1}{\lambda_i-u_j}$  $\square$.

Here we present a more explicit form of the functions $f$ and $g$.
Let $S(\lambda)=(\lambda-\lambda_1)...(\lambda-\lambda_{n+1-m}),$ where   $\lambda_i$ are pairwise distinct constants and $R(\lambda)=(\lambda-u_1)...(\lambda-u_n)$. Functions $f,~g$ 
can be written as 
\begin{equation} \label{pot2} 
f(u_1,...,u_{n+1-m},v_1,...,v_{m},\lambda)=
\end{equation} 
$$\frac{S(\lambda)\sum_{1\leq i_1<...<i_{m+1}\leq n+1-m}\phi_{i_1...i_{m+1}}(\lambda-\lambda_{i_1})...(\lambda-\lambda_{i_{m+1}})R(\lambda_{i_1})^{-1}...R(\lambda_{i_{m+1}})^{-1}}
{R(\lambda)\sum_{1\leq i_1<...<i_{m+2}\leq n+1-m}\Delta_{i_1...i_{m+2}}R(\lambda_{i_1})^{-1}...R(\lambda_{i_{m+2}})^{-1}}$$
$$g(u_1,...,u_{n+1-m},v_1,...,v_{m},\lambda)=$$
$$\frac{S(\lambda)\sum_{1\leq i_1<...<i_{m+1}\leq n+1-m}\psi_{i_1...i_{m+1}}(\lambda-\lambda_{i_1})...(\lambda-\lambda_{i_{m+1}})R(\lambda_{i_1})^{-1}...R(\lambda_{i_{m+1}})^{-1}}
{R(\lambda)\sum_{1\leq i_1<...<i_{m+2}\leq n+1-m}\Delta_{i_1...i_{m+2}}R(\lambda_{i_1})^{-1}...R(\lambda_{i_{m+2}})^{-1}}$$
where 
$$\phi_{i_1...i_{m+1}}=\prod_{1\leq \alpha<\beta\leq m+1}(\lambda_{i_{\alpha}}-\lambda_{i_{\beta}})\det\left(
\begin{array}{ccc}
a_{i_1} & ... & a_{i_{m+1}}  \\
c_{i_1,1} &...& c_{i_{m+1},1}\\
...&...&...\\c_{i_1,m}&...&c_{i_{m+1},m}   \end{array}
\right),~~~$$$$\psi_{i_1...i_{m+1}}=\prod_{1\leq \alpha<\beta\leq m+1}(\lambda_{i_{\alpha}}-\lambda_{i_{\beta}})\det\left(
\begin{array}{ccc}
b_{i_1} & ... & b_{i_{m+1}}  \\
c_{i_1,1} &...& c_{i_{m+1},1}\\
...&...&...\\c_{i_1,m}&...&c_{i_{m+1},m}   \end{array}
\right),$$ 
$$\Delta_{i_1...i_{m+2}}=\prod_{1\leq \alpha<\beta\leq m+2}(\lambda_{i_{\alpha}}-\lambda_{i_{\beta}})\det\left(
\begin{array}{ccc}
a_{i_1} & ... & a_{i_{m+2}}  \\
b_{i_1} & ... & b_{i_{m+2}}  \\
c_{i_1,1} &...& c_{i_{m+2},1}\\
...&...&...\\c_{i_1,m}&...&c_{i_{m+2},m}   \end{array}
\right),$$
and $a_i,~b_i,~\lambda_i,~c_{ij}$ are constants. In particular, if $m=0$, we have
\begin{equation} \label{pot1} 
f(u_1,...,u_{n+1},\lambda)=\frac{S(\lambda)\sum_{i=1}^{n+1}(\lambda-\lambda_i)a_iR(\lambda_i)^{-1}}{R(\lambda)\sum_{1\leq i<j\leq n+1}(\lambda_i-\lambda_j)(a_ib_j-a_jb_i)R(\lambda_i)^{-1}R(\lambda_j)^{-1}}
\end{equation} 
$$g(u_1,...,u_{n+1},\lambda)=\frac{S(\lambda)\sum_{i=1}^{n+1}(\lambda-\lambda_i)b_iR(\lambda_i)^{-1}}{R(\lambda)\sum_{1\leq i<j\leq n+1}(\lambda_i-\lambda_j)(a_ib_j-a_jb_i)R(\lambda_i)^{-1}R(\lambda_j)^{-1}}$$
Note that this form is useful for changing coordinates. For example, we can choose $R(\lambda)=v_1+v_2\lambda+...+v_{n}\lambda^{n-1}+\lambda^n$.

\subsection{Degenerations}

 The construction of Proposition 1 admits a limit when some of $\lambda_i$ coincide. All formulas are valid in this case except (\ref{qs}). The formula (\ref{qs}) should be rewritten as 
\begin{equation} \label{qs1}
q_{n-m-3+j,i}(u)=\frac{\bar{c}_{1,j}+\bar{c}_{2,j}u+...+\bar{c}_{n+1-m,j}u^{n-m}}{(u-\lambda_1)...(u-\lambda_{n+1-m})},
\end{equation}
where $\bar{c}_{i,j}$ are constants. In (\ref{qs1}) some of $\lambda_i$ may coincide. An analog of (\ref{qs}) can be obtained from (\ref{qs1}) by the partial fraction expansion of rational functions in $u$. Note that if 
$\lambda_i=\lambda_{i+1}=...=\lambda_{i+k}$, then the corresponding densities of conservation laws (\ref{cons1}) read 
$$\mu_{i,j}=\frac{\phi_1}{(\lambda_i-u_1)^j}+...+\frac{\phi_n}{(\lambda_i-u_n)^j}, \qquad \nu_{i,j}=\frac{\psi_1}{(\lambda_i-u_1)^j}+...+\frac{\psi_n}{(\lambda_i-u_n)^j},$$
where $~j=1,...,k+1$.

 It is clear from (\ref{lax}) that  functions $f,~g$ of the form (\ref{fgrep}) with degrees of 
polynomials $P,~Q,~R,~S$ equal $m+1,~m+1,~n,~n+1-m$ respectively   transform  as $$f\to\bar{f}=\frac{(c\lambda+d)^2S(\bar{\lambda})P(\bar{\lambda})}{R(\bar{\lambda})}, \qquad
g\to\bar{g}=\frac{(c\lambda+d)^2S(\bar{\lambda})Q(\bar{\lambda})}{R(\bar{\lambda})}$$
 under transformations (\ref{frac}).
 In particular, one of the values of $\lambda_i$ can be sent to infinity. If $\lambda_i=\infty$ and this root has multiplicity $k$, 
say $\lambda_{n+1-m}=...=
\lambda_{n+2-m-k}=\infty$, then $S(\lambda)=(\lambda-\lambda_1)...(\lambda-\lambda_{n+1-m-k})$ and (\ref{qs1}) takes the form 
$$q_{n-m-3+j,i}(u)=\frac{\bar{c}_{1,j}+\bar{c}_{2,j}u+...+\bar{c}_{n+1-m,j}u^{n-m}}{(u-\lambda_1)...(u-\lambda_{n+1-m-k})}.$$

Consider the case when all $\lambda_1=...=\lambda_{n+1-m}=\infty$. We have $S(\lambda)=1$ and 
$$q_{n-m-3+j,i}(u)=\bar{c}_{1,j}+\bar{c}_{2,j}u+...+\bar{c}_{n+1-m,j}u^{n-m}.$$
Since we can replace the equations in system (\ref{lins}) by any of their linear combinations, in particular we can subtract linear combinations of equations (\ref{uuu}) from 
other equations, we can assume without loss of generality that
$$q_{n-m-3+j,i}(u)=\bar{c}_{1,j}u^{n-m-2}+\bar{c}_{2,j}u^{n-m-1}+\bar{c}_{3,j}u^{n-m}.$$
Therefore we can only have $m=0$ or $m=1$. If $m=0$ and our constants $c_{i,j}$ are chosen in such a way that $$q_{n-2,i}(u)=u^{n-2},~q_{n-1,i}(u)=u^{n-1},$$
then our system takes the form
$$u_{i,t}=\sum_{j\ne i} u_j  \,\, u_{i,x}+ \prod_{j\ne i}(u_i-u_j)^{-1}, \qquad i=1,...,n.$$
This system has appeared in \cite{ferfor}, where a different problem was studied. If $n=2$, then this system coincides with (\ref{gibts}).

\subsection{General scheme}

 The main idea of our construction from Proposition 1 can be described as follows. Let $L(\lambda)$ be the difference between the  left and right hand sides of (\ref{comp}). If $f,~g$ are given by (\ref{fgrep}), 
then $L$ is a rational function in $\lambda$ and its numerator is the left hand side of (\ref{comp4}).
If the degrees of the polynomials $P,~Q,~R,~S$ are $m+1,~m+1,~n,~n+1-m$ 
respectively, then the degree of the numerator of $L$ equals $n+m+1$ and therefore the identity $L=0$ is equivalent to a system of  $n+m+2$ non-homogeneous hydrodynamic type equations. To provide $m+2$ linear relations between them we impose   constraints of the form 
$$c_{1,j}L(\lambda_1)+...+c_{n+1-m,j}L(\lambda_{n+1-m})=0, \qquad j=1,...,m+2,$$ 
where the $c_{i,j}$ are constants. Since $L(\lambda_i)=f(\lambda_i)_t-g(\lambda_i)_x$, these constraints follow from 
\begin{equation} \label{const}
 c_{1,j}f(\lambda_1)+...+c_{n+1-m,j}f(\lambda_{n+1-m})=a_j,\qquad c_{1,j}g(\lambda_1)+...+c_{n+1-m,j}g(\lambda_{n+1-m})=b_j,
\end{equation}
 where $j=1,...,m+2$ and $a_j, b_j$ are arbitrary constants.\footnote{Without loss of generality we can set $a_j=
\delta_{m+1,j},~b_j=\delta_{m,j}$.} The coefficients of $P$ and $Q$ are uniquely expressed from (\ref{const}) in term of coefficients of $R$.

 It is possible to generalize the construction described in Section 3.1 in the following way. We still assume that $f$ and $g$ are given by (\ref{fgrep}), where the degrees of polynomials $P,~Q,~R,~S$ are $m+1,~m+1,~n,~n+1-m$ and  $0\leq m\leq n-1.$  In the case $m<n-1$ we use representation (\ref{fg0}). If $m=n-1,$ then 
 \begin{equation} \label{fgn}
 f=S(\lambda)\left(\phi_0+\frac{\phi_1}{\lambda-u_1}+...+\frac{\phi_n}{\lambda-u_n}\right),\quad g=S(\lambda)\left(\psi_0+\frac{\psi_1}{\lambda-u_1}+...+\frac{\psi_n}{\lambda-u_n}\right).
\end{equation}
In our generalization we suppose that (\ref{const}) is valid for $j=1,...,m+2-k$. We look for the remaining $k$ 
linear constraints   in the form 
\begin{equation} \begin{array}{c}\label{constgen} d_{1,j,i}f(\lambda_i)+d_{2,j,i}f^{\prime}(\lambda_i)+...+d_{l,j,i}\frac{f^{(l-1)}(\lambda_i)}{(l-1)!}=0, \\[4mm] d_{1,j,i}g(\lambda_i)+d_{2,j,i}g^{\prime}(\lambda_i)+...+d_{l,j,i}\frac{g^{(l-1)}(\lambda_i)}{(l-1)!}=0,
\end{array}
\end{equation}
where $j=1,...,k_i,~i=1,...,n+1-m$ and $k_1+...+k_{n+1-m}=k$. The coefficients $d_{s.j,i}$ have to be found from the identity  
\begin{equation} \label{constL}
 d_{1,j,i}L(\lambda_i)+d_{2,j,i}L^{\prime}(\lambda_i)+...+d_{l,j,i}\frac{L^{(l-1)}(\lambda_i)}{(l-1)!}=0.
\end{equation}
Let $$f=(\lambda-\lambda_i)f_0+(\lambda-\lambda_i)^2f_1+..., \qquad g=(\lambda-\lambda_i)g_0+(\lambda-\lambda_i)^2g_1+...,$$$$L=L_0+(\lambda-\lambda_i)L_1+(\lambda-\lambda_i)^2L_2+... \,.$$ We 
omit the index $i$ in the coefficients of these Taylor expansions for simplicity. 
Substituting the Taylor expansions into (\ref{comp}),  we obtain
$$L_0=f_{0,t}-g_{0,x},$$ 
$$L_i=f_{i,t}-g_{i,x}+\sum_{0\leq k < \frac{i}{2}}(i-2k)(f_kg_{i-k}-f_{i-k}g_k), \qquad i>0.$$
Substituting these expressions into (\ref{constL}), we see that the terms $f_{j,t},~g_{j,x}$ cancel out by virtue of (\ref{constgen}). The remaining parts of the expressions are bilinear in $f_j,~g_j$. The vanishing of such parts leads to constraints for the coefficients $d_{s,j,i}$. Given a set of coefficients $d_{s,j,i}$ satisfying these constraints, we substitute expressions (\ref{fg0}) for $f,~g$ into 
(\ref{constgen}) and obtain a set of $k=k_1+...+k_{n+1-m}$ linear equations for $\phi_i,~\psi_i$. We combine this system of linear equations with (\ref{uuu}) and (\ref{lind}), where $\mu_i,~\nu_i$ are given by (\ref{munu}) and 
$j=1,...,m+2-k$. The whole set of linear relations guarantees that the identity $L=0$ is equivalent to a system of the form (\ref{geninhom}).

We do not describe here all admissible sets of coefficients $d_{s,j,i}$ in (\ref{constgen}) but just discuss two admissible cases.

{\bf Case 1.} Suppose that  (\ref{constgen}) takes a form $f_s=d_{s,i}f_0,~g_s=d_{s,i}g_0,~s=1,...,k_i.$ 
Then we have also $L_s=d_{s,i}L_0,~s=1,...,k_i.$ 

Let $m<n-1$. In this case $f,~g$ are given by (\ref{fg0}), where $\phi_1,...,\phi_n,\psi_1,...,\psi_n$ are defined as the solution of the linear system combined by the following three parts:

Part 1. The system (\ref{uuu}) with $j=0,...,n-m-3$. This system is empty if $m=n-2$.

Part 2. The system (\ref{lind}), where $\mu_i,~\nu_i$ are given by (\ref{munu}) and $j=1,...,m+2-k$. 

Part 3. The system 
\begin{equation} \label{fipsi}
\left(\frac{1}{(\lambda_i-u_1)^{j+1}}-\frac{p_{j,i}}{\lambda_i-u_1}\right)\phi_1+...+\left(\frac{1}{(\lambda_i-u_n)^{j+1}}-\frac{p_{j,i}}{\lambda_i-u_n}\right)\phi_n=0, 
\end{equation}
$$\left(\frac{1}{(\lambda_i-u_1)^{j+1}}-\frac{p_{j,i}}{\lambda_i-u_1}\right)\psi_1+...+\left(\frac{1}{(\lambda_i-u_n)^{j+1}}-\frac{p_{j,i}}{\lambda_i-u_n}\right)\psi_n=0,$$
where $j=1,...,k_i,~i=1,...,n+1-m$ and $k_1+...+k_{n+1-m}=k$.

If $m=n-1$ then we have to take $k=n-1$. The functions $f,~g$ are given by (\ref{fgn}), where $\phi_0,...,\phi_n,\psi_0,...,\psi_n$ is the solution of the linear system combined by the following two parts:

Part 1. The system (\ref{lind}), where $\mu_i,~\nu_i$ are given by 
$$\mu_i=\phi_0+\frac{\phi_1}{\lambda_i-u_1}+...+\frac{\phi_n}{\lambda_i-u_n},\qquad \nu_i=\psi_0+\frac{\psi_1}{\lambda_i-u_1}+...+\frac{\psi_n}{\lambda_i-u_n}$$
and $j=1,2$. 

Part 2. The system 
\begin{equation} \label{fipsi1}
-p_{j,i}\phi_0+\left(\frac{1}{(\lambda_i-u_1)^{j+1}}-\frac{p_{j,i}}{\lambda_i-u_1}\right)\phi_1+...+\left(\frac{1}{(\lambda_i-u_n)^{j+1}}-\frac{p_{j,i}}{\lambda_i-u_n}\right)\phi_n=0, 
\end{equation}
$$-p_{j,i}\psi_0+\left(\frac{1}{(\lambda_i-u_1)^{j+1}}-\frac{p_{j,i}}{\lambda_i-u_1}\right)\psi_1+...+\left(\frac{1}{(\lambda_i-u_n)^{j+1}}-\frac{p_{j,i}}{\lambda_i-u_n}\right)\psi_n=0,$$
where $j=1,...,k_i,~i=1,2$ and $k_1+k_2=n-1$. $\square$

{\bf Case 2.} Suppose that (\ref{constgen}) takes the form 
\begin{equation} \label{lincon}
f_s=a_{s,i}f_0+b_{s,i}f_1, \qquad g_s=a_{s,i}g_0+b_{s,i}g_1, \qquad s=2,...,k_i,
\end{equation} 
where the coefficients $a_{i,j}$ and $b_{p,q}$ satisfy the following equations 
$$(s-2)a_{s-1,i}+b_{s,i}+\sum_{2\leq k<\frac{s}{2}}(s-2k)(b_{k,i}a_{s-k,i}-b_{s-k,i}a_{k,i})=0, \qquad s=2,3,....$$
In particular, $b_{2,i}=0.$ Then we have also $L_s=a_{s,i}L_0+b_{s,i}L_1,~s=2,...,k_i.$ The explicit form of a linear system for $\phi_1,...,\phi_n,\psi_1,...,\psi_n,$ where $m<n-1$ (resp. for $\phi_0,...,\phi_n,\psi_0,...,\psi_n,$ where $m=n-1$) 
can be obtained straightforwardly by substitution of (\ref{fg0}) (resp. (\ref{fgn})) into (\ref{lincon}) and combining these equations with (\ref{uuu}) and (\ref{lind}). $\square$

{\bf Remark 3.} For any admissible coefficients $d_{s,j,i}$, if  $\lambda_i$ are distinct, then we must have
\begin{equation} \label{nmk}
2m+1-n\leq k\leq m, \qquad k_1+...+k_{n+1-m}=k. 
\end{equation}
Indeed, we must have at least two equations of the form (\ref{lind}) for $\phi_i$ (resp. for $\psi_i$), so $k\leq m$. On the 
other hand, the rank of the matrix $(c_{j,i})$ should be equal to $m+2-k$. Therefore, $m+2-k\leq n+1-m$ or $2m+1\leq n+k$.
In particular, if $m=n-1$, then $k=n-1$. $\square$

Consider the case $n=2$. It is easy to see that (\ref{nmk}) admits only two solutions:

1. $m=0,~k=k_1=k_2=k_3=0$. This is the case of Proposition 1. Explicit formulas are given by (\ref{Ex1}).

2. $m=1,~k=1$. Without loss of generality we can set $k_1=1,~k_2=0$. There are only two possibilities for coefficients $d_{s,j,i}$ and both are described by the cases 1 and 2 above. In case 1 we have
$f_1=af_0,~g_1=ag_0$ where $a$ is an arbitrary constant and $f_i,~g_i$ are coefficients of the Taylor expansion of $f,~g$ respectively at $\lambda=\lambda_1$. Explicit formulas are given by (\ref{Ex3}). In case 2 we have
$f_2=af_0,~g_2=ag_0,$ where $a$ is an arbitrary constant and $f_i,~g_i$ are coefficients of the Taylor expansion of $f,~g$ respectively at $\lambda=\lambda_1$. Explicit formulas are given by (\ref{Ex2}).

\section{Classification in the case $n=2$}

Consider the case $n=2.$ In this case we do not assume that $f$ and $g$ are rational functions in $\lambda$. Denote $u_1=u,~u_2=v$. Expanding the functional equation (\ref{funeqgen}) a neighborhood of the diagonal $v=u,$ 
we obtain $q_{1,1}(u)=q_{1,2}(u)$ and
$h_1(\lambda,u)=h_2(\lambda,u)$. Let $$q_{1,1}(u)=q_{1,2}(u)=\frac{1}{a(u)},\qquad  h_1(\lambda,u)=h_2(\lambda,u)=\frac{h(\lambda,u)}{a(u)}.$$ 
Then system (\ref{geninhom}) takes the form
\begin{equation}   \label{sysnorm}
u_t=a(v)u_x+h(u,v), \qquad v_t=a(u)v_x+h(v,u) 
\end{equation}
and the corresponding Lax pair (\ref{lax}) is defined by
\begin{equation}   \label{fg}
f(u,v,\lambda)=\frac{h(\lambda,u)-h(\lambda,v)}{a(u)-a(v)}, \qquad g(u,v,\lambda)=\frac{a(v)h(\lambda,u)-a(u)h(\lambda,v)}{a(u)-a(v)}.
\end{equation}
Here $h$ is a function with a simple pole on the diagonal. Formula (\ref{funeqgen}) yields the following functional equation for this function:
\begin{equation} \label{funeq1}
h(\lambda,v)h(\lambda,u)_{\lambda}-h(\lambda,u)h(\lambda,v)_{\lambda}+h(u,v)h(\lambda,u)_u-h(v,u)h(\lambda,v)_v
-$$$$\frac{h(u,v)a^{\prime}(u)-h(v,u)a^{\prime}(v)}{a(u)-a(v)}(h(\lambda,u)-h(\lambda,v))=0.
\end{equation}

{\bf Remark 4.} Suppose (\ref{funeq1}) holds and $h$ has a simple pole on the diagonal. Then the following  Gibbons-Tsarev type  system with two fields $u,~v$ \cite{Gibt1, ferhus1, odsok2} is compatible:
\begin{equation} \label{gt}
\partial_i p_j=h(p_j,p_i)\partial_i u,\quad \partial_i v=a(p_i)\partial_i u,\quad \partial_i\partial_j u=\frac{h(p_i,p_j)a^{\prime}(p_i)-h(p_j,p_i)a^{\prime}(p_j)}{a(p_i)-a(p_j)}\partial_i u\partial_j u. 
\end{equation}
Here $i\ne j=1,...,N$; $p_1,...,p_N,u$ are functions in $r_1,...,r_N$, $\partial_i=\frac{\partial}{\partial r_i}$ and $N$ is arbitrary. Note that there exist Gibbons-Tsarev type systems that 
have a slightly different structure. $\square$

{\bf Proposition 2.} For each Gibbons-Tsarev type  system of the form (\ref{gt}) one can construct a non-homogeneous system (\ref{diaginhom}) with $n=2$ possessing a Lax representation. 

{\bf Proof.} Set $N=2$ in (\ref{gt}) and consider $p_1,~p_2$ as functions of $u,~v$. We have 
$$\partial_i p_j=(p_j)_u\partial_i u+(p_j)_v\partial_i v=(p_j)_u\partial_i u+(p_j)_va(p_i)\partial_i u=h(p_j,p_i)\partial_i u$$ or $(p_j)_u+(p_j)_va(p_i)=h(p_j,p_i),$ where $i\ne j=1,2$.
Moreover, it is known \cite{odsok2} that each Gibbons-Tsarev type  system admits so-called dispersionless Lax operator. It is a function $L(\lambda,r_1,...,r_N)$ defined (up to transformations of the form $L\to q(L)$) by 
the following system
$$\partial_i L=h(\lambda,p_i)L_{\lambda}\partial_i u, \qquad i=1,...,N.$$
Note that this system is compatible by virtue of (\ref{gt}). Taking $N=2,$ we may consider $L$ as a function of $u,~v,~\lambda$. As above we obtain $L_u+L_va(p_i)=h(\lambda,p_i)L_{\lambda},~i=1,2$ or 
$L_u=f(p_1,p_2,\lambda)L_{\lambda},~L_v=g(p_1,p_2,\lambda)L_{\lambda},$ where $$f=\frac{a(p_2)h(\lambda,p_1)-a(p_1)h(\lambda,p_2)}{a(p_2)-a(p_1)}, \qquad g=\frac{h(\lambda,p_1)-h(\lambda,p_2)}{a(p_1)-a(p_2)}. \qquad \square$$

{\bf Remark 5.} In \cite{ferfor} several systems of the form (\ref{sysnorm}) possessing a conservation law with density and flux that depend on $u,v,u_x,v_x$  were found. The existence of 
such a conservation law was proposed as an indication of complete integrability. Most of the systems from \cite{ferfor} do not satisfy our functional equation (\ref{funeq1}). $\square$

Let us present several solutions of the functional equation (\ref{funeq1}).

{\bf Example 1.} The functions 
\begin{equation}\label{Ex1}
h(u,v)=\frac{S_3(u)\, S_3(v)}{W_2(v)(u-v)}, \qquad a(u)=\frac{P_2(u)}{W_2(u)}.
\end{equation}
where  $S_3,$  $P_2$ and  $W_2$ are arbitrary polynomials of degree 3, 2 and 2 correspondingly, satisfy (\ref{funeq1}).
For equation (\ref{gibts}) we have $S_3(u)=1,$ $ W_2(u)=-1,$ $ P_2(u)=-u.$ $\square$

{\bf Example 2.} Let 
\begin{equation}\label{Ex2}
h(u,v)=\frac{Q(u)\, Q(v) P(u) \Big(P(v)^2 P(u)+(u-v) Q(v)\Big)}{R(v)(u-v)}, \qquad a(u)=\frac{S(u)}{R(u)},
\end{equation}
where $P(u)=p_1 u+p_0$, $Q(u)=q_1 u+q_0$ are arbitrary polynomials and the coefficients of polynomials $S(u)=s_2 u^2+s_1 u+s_0$ and $R(u)=r_2 u^2+r_1 u+r_0$ satisfy 
the same linear equation 
\begin{equation} \label{rel}
2 p_0 q_0 x_2-(p_0 q_1 +p_1 q_0) x_1+2 p_1 q_1 x_0=0.
\end{equation}
Then the functions $h,a$ satisfy (\ref{funeq1}).
Notice that (\ref{rel}) means that the double ratio $(pq,z_1z_2)$ equals $-1$, where $p,q,z_1,z_2$ are roots of the polynomials 
$P(u),Q(u)$ and $x_2 u^2+x_1 u+x_0$ respectively.  $\square$

{\bf Example 3.} The functions
\begin{equation}\label{Ex3}
h(u,v)=\frac{Q(u)\, Q(v) R(u) \Big(R(v) R(u)+k (u-v)\Big)}{T(v)(u-v)}, \qquad a(u)=\frac{S(u)}{T(u)},
\end{equation}
where  $Q(u)=q_1 u+q_0$ $S(u)=s_1 u+s_0$, $R(u)=r_1 u+r_0,$ $T(u)=t_1 u+t_0$ are arbitrary polynomials, satisfy (\ref{funeq1}).   $\square$

It is easy to verify that the classes of solutions described in Examples 1-3 are invariant with respect to the  transformations
\begin{equation} \label{tran1}
u\to \frac{k_1 u+k_2}{k_3 u+k_4}, \qquad v\to \frac{k_1 v+k_2}{k_3 v+k_4}
\end{equation}
and 
\begin{equation} \label{tran2}
x\to r_1 x+r_2 t, \qquad t\to r_3 x+r_4 t,
\end{equation}
where $k_i$ and $r_i$ are arbitrary constants. 

{\bf Remark 6.} Using transformations (\ref{tran1}), (\ref{tran2}), we can bring some of polynomials from Examples 1-3 to a canonical form. For instance, in the generic case of Example 1 one can bring 
polynomial $S_3$  to $S_3(u)=u(u-1)$. $\square$

{\bf Theorem 1.} Any solution of (\ref{funeq1}) that has a simple pole on the diagonal is given by (\ref{Ex1}), (\ref{Ex2}) or (\ref{Ex3}) up to transformations (\ref{point}).

{\bf Proof.} Let 
\begin{equation} \label{series}
h(z,x)=\frac{a_{-1}(x)}{z-x}+a_0(x)+a_1(x)(z-x)+....
\end{equation}
Relations (\ref{sysnorm}),  (\ref{fg}), (\ref{funeq1}) admit arbitrary transformations of the form 
\begin{equation}\label{point}
u \to \phi(u), \qquad v \to \phi(v).
\end{equation}
Normalizing $a(u)$ by $u$  with the help of (\ref{point}) and expanding our functional equation at $\lambda=u,$ we obtain
\begin{equation} \label{PDE1}
 h(u,v)_u=\frac{1}{2a_{-1}(u)(u-v)}\Big(a_{-1}(u)h(v,u)+((u-v)a_{-1}^{\prime}(u)-a_{-1}(u))h(u,v)\Big).
\end{equation}
For the function
\begin{equation} \label{gg}
q(u,v)= \frac{h(u,v)}{\sqrt{a_{-1}(u)a_{-1}(v)}}
\end{equation} 
equation (\ref{PDE1}) takes the form
\begin{equation} \label{PDE2}
q(u,v)_u=\frac{q(v,u)-q(u,v)}{2(u-v)}.
\end{equation}
From (\ref{PDE2}) it follows that $q(u,v)_u=q(v,u)_v$. Differentiating (\ref{PDE2}) by $v$ and eliminating $q(v,u)$ and  $q(v,u)_v,$ we arrive at the following Euler-Darboux equation 
\begin{equation} \label{darbu}
q_{u v}=\frac{3}{2} \frac{q_u}{u-v}-\frac{1}{2} \frac{q_v}{u-v}
\end{equation}
for $q(u,v).$

We are interested in solutions of (\ref{darbu}) of the form $q(u,v)=\frac{1}{u-v}+G(u,v)$, where $G(u,v)$ is holomorphic on the diagonal $u=v$. It is easy to verify that for any such a solution the function $G$ is of the form
$$
G(u,v)= \sum_{0}^{\infty} \frac{s(u+v)^{(2k)}}{(k!)^2 2^{2k}} (u-v)^{2 k}-\sum_{1}^{\infty} \frac{s(u+v)^{(2k-1)}}{k! (k-1)! 2^{2k-1}} (u-v)^{2 k-1}
$$
for some function $s(x)$. The functions $a_{-1}(x),a_0(x)$ from (\ref{funeq1}) and $s(x)$ are related as follows
\begin{equation} \label{a0}
a_0=\frac{a_{-1}'}{2}+s a_{-1}.
\end{equation}

We will show that for any solution (\ref{series}) of (\ref{funeq1}) the corresponding function $s$ is rational (in contrast with $a_{-1}, a_0$). For any rational $s(x)$
the corresponding function $G$ can be easily reconstructed in closed form (cf. \cite{marsok}). In particular, if 
$s(x)=1/(x-k)$ then $G(u,v)=T(u,v,k),$ where 
\begin{equation} \label{TT}
T(u,v,k) =\frac{2}{(\sqrt{u-k})+\sqrt{v-k}) \sqrt{v-k}}.
\end{equation}
The multiple pole  $s(x)=1/(x-k)^2$ corresponds to $\displaystyle \frac{\partial T(u,v,k)}{\partial k}$ and so on.

Substituting (\ref{series}), (\ref{a0}) into (\ref{funeq1}) and expanding in a small neighborhood of $u=v=z$, we find that all coefficients $a_i$, with $i>0$ are uniquely determined through 
$a_{-1}$ and $s.$ These two functions satisfy an overdetermined system of ODEs. Eliminating $a_{-1},$ we arrive at  another system for $s$ only. This system contains two ODEs of fifth and fourth orders. 
Differentiating the fourth order ODE and eliminating the fifth derivative by fifth order equation, we get several more fourth order ODEs. The final system is so complicated that usual algorithms of computer algebra 
such as differential Groebner basis technique certainly do not work. Fortunately, the system admits a group of point symmetries. 

Although we fix $a(u)=u$, the equation (\ref{funeq1}) still admits symmetries. Namely, after any transformation of the form (\ref{tran2}) the function $a(u)$ becomes fractional-linear and we can bring it back to $u$ by 
an appropriate transformation (\ref{tran1}).
The existence of this $GL(2)$-action implies the symmetry group
$$
s(u) \to \frac{k_3}{2 (k_3 u+k_4)}+\frac{\displaystyle (k_1 k_4-k_2 k_3)\, s\Big(\frac{k_1 u+k_2}{k_3 u+k_4}\Big)}{(k_3 u+k_4)^2}
$$
of the ODE system for $s(u)$.
Taking the simplest differential invariants 
$$
x=\frac{s''+12 s s'+16 s^3}{(s'+2 s^2)^{3/2}}, \qquad y=\frac{s'''+24 s s''+144 s^2 s'+144 s^4}{(s'+2 s^2)^{2}}
$$
of this action for new dependent and independent variables, we can reduce the order of equations by 3. The equations of fourth order turn into very complicated first order equations for  the function $y=G(x)$.  
Considering these equations as polynomials in $x, G$ and $G'$ and eliminating  $G'$   by using the resultant several times, we find that the whole system is equivalent to the following single algebraic equation
$$
G^6+288\, G^5-216\, (5 x^2-124)\, G^4+54\, (13 x^4-944 x^2+16768)\, G^3
$$
$$
-3888\,(27 x^4+40 x^2-3632)\, G^2+5832\,(63 x^6-1124 x^4+1856 x^2+17920)\,G
$$
$$
-729\,(343 x^8-9376 x^6+69120 x^4-118784 x^2-409600)=0.
$$
The general solution of this third order ODE for $s(u)$ is given by 
\begin{equation} \label{case1}
s(u)=\frac{1}{4 (u-k_1)}+\frac{1}{4 (u-k_2)}+\frac{k_3}{(u-k_1)^2},
\end{equation} 
 where $k_i$ are arbitrary constants. Besides (\ref{case1}) there is a special solution of the form 
 \begin{equation} \label{case2}
s(u)=\frac{1}{4 (u-k_1)}+k_2.
\end{equation} 

The next step is to find the function $a_{-1}$ from the initial ODE system for $s(u)$ and $a_{-1}(u).$ It turns out that if $s$ is given by (\ref{case1}) with $k_3=0$ then this system is equivalent 
to a single non-linear ODE of fourth order for $a_{-1}$. It can be linearized by the substitution $a_{-1}=(u-k_1)(u-k_2) P(u)^2.$ Solving the linear ODE for $P$, we find
$$
a_{-1}(u)=(u-k_1)(u-k_2) \Big(P_1(u) \sqrt{u-k_1}+P_2(u) \sqrt{u-k_2} \Big)^2,
$$
where $P_i$ are arbitrary first degree polynomials. The function $g(u,v)$ is reconstructed from $s$ as follows
$$
g(u,v)=\frac{1}{u-v}+\frac{1}{4} T(u,v,k_1)+\frac{1}{4} T(u,v,k_2),
$$
where $T(u,v,k)$ is given by (\ref{TT}).
The function $h(u,v)$ is defined by (\ref{gg}). The radicals $\sqrt{u-k_1}$ and $\sqrt{u-k_2}$ can be removed by an appropriate transformation of the form (\ref{point}) and as 
result we get the solution described in Example 1. 

Example 2 corresponds to  (\ref{case1}) with $k_3\ne 0$. In this case we have 
$$
g(u,v)=\frac{1}{u-v}+\frac{1}{4} T(u,v,k_1)+\frac{1}{4} T(u,v,k_2)+k_3  \frac{\partial T(u,v,k_1)}{\partial k_1}
$$
and $
a_{-1}=\alpha (u-k_1)^3 (u-k_2)^2,
$
where $\alpha$ is arbitrary constant.

The degeneration $k_2=k_1$, $k_3\ne 0$ of (\ref{case1}) gives rise to  Example 3. In this case $a_1=(u-k_1)^3 P_1^2,$ where $P_1$ 
is an arbitrary first degree polynomial. 

Solutions of (\ref{funeq1}) corresponding to other degenerations can be written in one of the forms (\ref{Ex1}), (\ref{Ex2}) or (\ref{Ex3}) with special polynomials there. In particular, 
the degeneration  $k_2=k_1$, $k_3=0$ corresponds to (\ref{Ex1}), where the polynomials $P_2$ and $Q_2$ have a common root. $\square$

\section{Examples in the case $n=3$}

In the case $N=3$ integrable systems have the form 
\begin{equation}   \label{gengibtsN3}
\begin{array}{c}
u_t=a_1(v,w) u_x+b_1(u,v,w), \\[3mm] v_t=a_2(u,w) v_x+b_2(u,v,w), \\[3mm] w_t=a_3(u,v) w_x+b_3(u,v,w),
\end{array}
\end{equation}
where
$$
a_1(u,v)=a_2(u,v)=a_3(u,v)=\frac{B(u)-B(v)}{A(u)-A(v)},
$$
$$
b_1(u,v,w)=\frac{X(v,u)-X(w,u)}{A(v)-A(w)}, \qquad b_2(u,v,w)=\frac{X(w,v)-X(u,v)}{A(w)-A(u)}, $$$$ b_3(u,v,w)=\frac{X(u,w)-X(v,w)}{A(u)-A(v)}.
$$
The Lax representation  of the form (\ref{lax}) is given by
$$
f(u,v,w,\lambda)=\frac{X(u,\lambda)(A(v)-A(w))+X(v,\lambda)(A(w)-A(u))+X(w,\lambda)(A(u)-A(v))}{B(u)(A(v)-A(w))+B(v)(A(w)-A(u))+B(w)(A(u)-A(v))},
$$
$$
g(u,v,w,\lambda)=\frac{X(u,\lambda)(B(v)-B(w))+X(v,\lambda)(B(w)-B(u))+X(w,\lambda)(B(u)-B(v))}{B(u)(A(v)-A(w))+B(v)(A(w)-A(u))+B(w)(A(u)-A(v))}.
$$
In the above formulas the functions $X(u,\lambda),A(u),B(u)$ satisfy a complicated functional equation, which follows from (\ref{funeqgen}).

{\bf Remark 7.} Taking into account transformations (\ref{tran2}), we see that the following group of affine transformations is admissible: 
$$
A\to c_1 A+c_2 B+c_3, \qquad B\to c_4 A+c_5 B+c_6,
$$

\subsection{A class of solutions}

Let us describe explicitly all integrable equations with the function $X$ of the form 
\begin{equation}\label{anza}
X(u,\lambda)=\frac{R(u) S(\lambda)}{\lambda-u}
\end{equation}
for some functions $R,S.$ In this section we do not assume that  the functions $R$ and $S$ are polynomials. 
Note that the transformation of $X$ under (\ref{tran1}) is given by
\begin{equation} \label{ttran1}
R(u)\to (k_3 u+k_4) R\Big(\frac{k_1 u+k_2}{k_3 u+k_4} \Big), \qquad S(\lambda)\to (k_3 \lambda+k_4)^3 S\Big(\frac{k_1 \lambda+k_2}{k_3 \lambda+k_4} \Big).
\end{equation}
It can be straightforwardly verified that all examples found below belong to the class of equations described in Section 3. 

From  (\ref{funeqgen}) it follows that the functions $R,S$ satisfy several ODEs  linear in $S.$ The simplest of them are
$$
6 R R'' S^{(5)}+5 (3 R'R''-R R''') S^{(4)} =0,
$$
\begin{equation} \label{ssys}
R^2 S^{(6)}+12 R R' S^{(5)}+15 (2 R'^2-R R'') S^{(4)} =0,
\end{equation}
$$
R R'' S^{(6)}-8 R R''' S^{(5)}+5(R R''''-2 R'R'''-3 R''^2) S^{(4)} =0,
$$
and
\begin{equation}  \label{lin1}
R^4 \,S^{(5)}+5 R^3 R' \,S^{(4)}-20 R^3 R'' \,S^{(3)}+10 R^2 (3 R' R''-R R''') \,S''- \end{equation}
$$
5 R (R^2 R^{(4)}-12 R R''^2+12 R'^2 R'') \,S'-(R^3 R^{(5)}-30 R^2 R'' R'''+90 R R' R''^2-60 R'^3 R'')\,S .
$$

There are two different possibilities: {\bf Case A:} $S''''=0$ and {\bf Case B:} $S''''\ne 0.$ In Case B the determinant of the system of linear equations (\ref{ssys}) for $S^{(4)},S^{(5)},S^{(6)}$ should be zero. This leads to a fourth order ODE for $R$, whose solution is $R=W_2/W_1$, where $W_2$ and $W_1$ are arbitrary polynomials of degree 2 and 1, correspondingly. Using a transformation of the form (\ref{ttran1}), we bring $W_1$ to 1. 

In Case A besides (\ref{lin1}) and $S^{(4)}=0$ we use one  more equation linear in $S$. This equation of order 3 in $S$ and order 4 in $R$ is rather complicated and we do not present it here. Differentiating these equations, we get several more linear equations. Eliminating $S$ and its derivatives from the system thus obtained, we arrive at an overdetermined system of non-linear ODEs for $R$.
Investigating the compatibility of latter system, we find all possible functions $R.$ Given $R$ the corresponding cubic polynomial $S$ can be easily found from (\ref{lin1}).

One of possible pairs $R,S$ in Case A is given by {\bf Case A-1}:
$$R=\frac{S_3}{W},\qquad S=S_3, $$ where $S_3$ and $W$ are arbitrary polynomials of degree 3 and 2 (cf. Example 1). 

It is possible to show that otherwise  we have {\bf Case A-2}: $R=1/W$ up to transformations of the form (\ref{ttran1}). Notice that the polynomial $W$ can be reduced by a linear transformation (\ref{ttran1}) to one of the following canonical forms: $W(x)=x(x-1)$, $W(x)=x^2,$  $W(x)=x,$ or  $W(x)=1.$ Using (\ref{lin1}), we find the following solutions:
 $$R(x)=\frac{1}{x(x-1)}, \qquad S(x)=c_1+c_2 x (x-1),$$
  $$R(x)=\frac{1}{x^2}, \qquad S(x)=c_1+c_2 x^2,$$
   $$R(x)=\frac{1}{x}, \qquad S(x)=c_1+c_2 x +c_3 x^2.$$
For $W=1$ see Case B.

It is easy to verify that in Case B the only solution of (\ref{ssys}),(\ref{lin1}) is $R=1, S=S_4,$ where $S_4$ is an arbitrary fourth degree polynomial. 

Now we should find the functions $A$ and $B$ for all above cases.
It can be verified that $A$ and $B$ satisfy the same fifth order linear equation whose coefficients are differential polynomials in $R,S.$  Hence all possible functions $A(u)$ form a vector space $V$ of dimension $\le 5$. According to Remark 7 all possible functions $B$ belong to $V$ and ${\rm dim}\, V$ has to be not less then 3. 

It turns out that in Case B we have
$$R=1, \qquad S=S_4, \qquad A=\frac{A_4}{S_4}, \qquad  B=\frac{B_4}{S_4},  $$  
where $S_4,A_4,B_4$ are arbitrary polynomials of fourth degree. In this case ${\rm dim}\, V=5.$

In Case A-2 we obtain that ${\rm dim}\, V=4.$ If ${\rm deg}\, W=2$ then a basis of $V$ is given by
$$
A_1=1, \qquad A_2= \frac{1}{W},  \qquad A_3= \frac{1}{S W},  \qquad A_2= \frac{u}{S W}.
$$
In the case  ${\rm deg}\, W<2$ a basis is $\displaystyle \frac{u^i}{S W}, \quad i=0,1,2,3$.
 
For Case A-1 (if the pair $R,S$ does not belong to Cases B and A-2), then the vector space $V$ is three-dimensional with a basis  
$\displaystyle \quad \frac{u^i}{W}, \quad i=0,1,2$ (cf. Example 1).
 
The following example has a structure more complicated then (\ref{anza}):
 
{\bf Example 4.}  $$X(u,z)=\frac{(u-a)^2 (z-a)(z-b)\Big((a+b) u z-2 b^2 z-2 a b u+b^2 (a+b)\Big)}{(u-z) u},$$ 
a basis of $V$ is $u,1,u^{-1}.$

\vskip.3cm \noindent {\bf Acknowledgments.} The authors thank  E.V. Ferapontov and I. Marshall for fruitful
discussions.  V.S. thanks MPIM and A.O. thanks IHES for
hospitality and financial support. V.S. was partially supported by
the RFBR grant 11-01-00341-a.

\end{document}